# A New "Non-energetic" Route to Complex Organic Molecules in Astrophysical Environments: The C + $H_2O$ → $H_2CO$ Solid-state Reaction

Alexey Potapov[1], Serge Krasnokutski[1], Cornelia Jäger[1] and Thomas Henning[2]

*[1]Laboratory Astrophysics Group of the Max Planck Institute for Astronomy at the Friedrich Schiller University Jena, Institute of Solid State Physics, Helmholtzweg 3, 07743 Jena, Germany, email: alexey.potapov@uni-jena.de*

*[2]Max Planck Institute for Astronomy, Königstuhl 17, D-69117 Heidelberg, Germany*

**Abstract**

The solid-state reaction C + $H_2O$ → $H_2CO$ was studied experimentally following the co-deposition of C atoms and $H_2O$ molecules at low temperatures. In spite of the reaction barrier and absence of energetic triggering, the reaction proceeds fast on the experimental timescale pointing to its quantum tunneling mechanism. This route to formaldehyde shows a new "non-energetic" pathway to complex organic and prebiotic molecules in astrophysical environments. Energetic processing of the produced ice by UV irradiation leads mainly to the destruction of $H_2CO$ and the formation of $CO_2$ challenging the role of energetic processing in the synthesis of complex organic molecules under astrophysically relevant conditions.

## 1. Introduction

Understanding the reaction pathways to complex organic and prebiotic molecules at the conditions relevant to various astrophysical environments, such as prestellar cores, protostars and planet-forming disks, will provide insight in the molecular diversity during the planet formation process (Jörgensen, Belloche, & Garrod 2020). Modern astrochemical reaction networks contain thousands of reactions (see, e.g., astrochemical databases such as KIDA (http://kida.astrophy.u-bordeaux.fr/) and UDFA (http://udfa.ajmarkwick.net/)). However, these networks are very sensitive to inclusions of new reactions. Even the inclusion of one new reaction may lead to a considerable change of the predicted abundances of many molecules in the solid state as was recently demonstrated by the addition of the C + $H_2$ → $CH_2$ reaction into a well-developed network (Simončič et al. 2020). Thus, systematic studies of astrochemically relevant reactions, particularly, involving simple species (as such species form at the beginning of reaction networks and may define their further evolution) are essential for the development of reliable astrochemical models.

Chemical reactions leading to the formation of molecules in cosmic environments can be divided into two groups, gas phase and grain surface reactions. Solid-state (or surface) reaction pathways at conditions relevant to interstellar and circumstellar environments lead to a greater complexity of molecular species up to prebiotic molecules, such as amino acids and nucleobases (Bernstein et al. 2002; Muñoz Caro et al. 2002; Holtom et al. 2005; Meinert et al. 2011; Nuevo, Milam, & Sandford 2012; Krasnokutski, Jäger, & Henning 2020; Ioppolo et al. 2021). In these reactions, the surface of grains may play an important role acting not only as a location for reactants, but also as a catalyst. Concerning the catalytic role of dust grains, we refer the reader to a few relevant studies (Hill & Nuth 2003; Saladino et al. 2005; Potapov et al. 2019; Potapov, Jäger, & Henning 2020a, 2020b) and a review (Potapov & McCoustra 2021).

$H_2O$ ice is the main constituent of cosmic molecular ices covering the surface of dust grains in cold regions (Whittet 2003). Atomic carbon is one of the most abundant carbonaceous species in various astrophysical environments (Schilke et al. 1995; Gerin et al. 1998; Henning & Salama 1998; Tanaka et al. 2011). Thus, the reaction between $H_2O$ and C is of fundamental interest and may considerably influence the evolution of astrochemical reaction networks. The high reactivity of C atoms is well-known thanks to theoretical and experimental studies (Henning & Krasnokutski 2019; Krasnokutski, et al. 2020; Qasim et al. 2020a; Qasim et al. 2020b).

The low-temperature reaction C + $H_2O$ was previously studied in the gas phase as well as in the solid state (Schreiner & Reisenauer 2006; Hickson et al. 2016). In the gas phase, a reactivity increase below 100 K was observed. As the rate of the C + $D_2O$ reaction was much lower compared to that of the C + $H_2O$ reaction, the increased reactivity was explained by the presence of quantum tunneling mechanism (Hickson, et al. 2016). In the solid state, the formation of formaldehyde ($H_2CO$) was detected, but assigned to the reactivity of excited C atoms in singlet states (Schreiner & Reisenauer 2006). It is well-known from laboratory experimental studies that reactions referred as "non-energetic" (typically, thermal atom addition reactions at low temperatures) may lead to high molecular complexity (Linnartz, Ioppolo, & Fedoseev 2015; Krasnokutski et al. 2017; Potapov et al. 2017; Ioppolo, et al. 2021). As energetic processing, such as UV irradiation and particle (ions, electrons) bombardment, of molecular ices may play both constructive and destructive roles (Öberg 2016), "non-energetic" reactions may present an alternative route to the formation of complex organic molecules in astrophysical environments.

The aim of the present study was to investigate the low-temperature solid-state reaction C + $H_2O$ → $H_2CO$ involving ground-state C($^3P_J$) atoms. The temperatures of the experiments, 10 – 80 K, correspond to a wide range of astrophysical environments, from interstellar clouds through hot cores and protostars to planet-forming disks and debris disks. In addition, the effect of UV irradiation on the ice produced by co-deposition of C and $H_2O$ reactants was studied.

## 2. Experimental Part

The experiments were performed in the INterStellar Ice Dust Experiment (INSIDE) setup presented in detail elsewhere (Potapov, Jäger, & Henning 2019). $H_2O$ ice and C atoms were deposited onto a KBr substrate at 10, 20, 50, and 80 K and a base pressure of a few $10^{-10}$ mbar through two independent gas lines placed at 45º angle to the substrate. C atoms were produced by an atomic source via thermal evaporation. The source is described in detail elsewhere (Krasnokutski & Huisken 2014). Briefly, it operates as follows. Carbon powder is placed inside a tantalum tube, which is resistively heated. Carbon powder dissolves into tantalum and diffuses to the outer surface of the tube walls, from which it evaporates in the form of single carbon atoms only. The source was connected to one of the flanges of the main chamber of INSIDE through a valve, which at open let the C-atoms flow enter the chamber and reach the substrate. As the binding energy C•$H_2O$ is about 3500 K, we expect that all C-atoms adsorb on the surface at the given experimental conditions.

The H$_2$O deposition time in all experiments was 40 minutes with a constant deposition rate for H$_2$O of about $3\times10^{15}$ molecules minute$^{-1}$. The number of deposited H$_2$O molecules was calculated from the vibrational stretching band at 3250 cm$^{-1}$ using the band strength of $2\times10^{-16}$ cm molecule$^{-1}$ (Hudgins et al. 1993). C atoms were deposited during 35 minutes (between the 4$^{th}$ and the 38$^{th}$ minute of the H$_2$O deposition). The flux of C atoms from the source was estimated elsewhere by means of scanning tunnelling microscopy on an Ag(111) substrate (Krasnokutski et al. 2019). The H$_2$O/C ratio of the deposited species in our experiments was estimated to be more than 10 (up to a few tens considering high uncertainty of the estimated flux of C-atoms).

In an additional experiment, D$_2$O ice and C atoms were deposited at 10 K. In addition, in order to study the direct contribution of cold C atoms in the C + H$_2$O reaction, a layered H$_2$O/Ar/C sample was produced at 10 K. We deposited about 10 monolayers of H$_2$O, then 10 monolayers of Ar, and then C atoms for 3 minutes. The procedure was repeated 10 times. Efficient diffusion of C atoms in Ar ice is well-known (Thompson, Dekock, & Weltner 1971), thus, after the C deposition, C atoms were expected to diffuse through Ar layers and interact with the pre-deposited top layers of H$_2$O ice.

Two series of experiments have been performed: (i) C + H$_2$O deposition and subsequent heating the samples and (ii) C + H$_2$O deposition, UV irradiation at the deposition temperature, and subsequent heating the samples. In the second series of experiments, after deposition the ices produced by co-deposition of C and H$_2$O reactants were irradiated for 2 hours at 45º incidence by a broadband deuterium lamp (L11798, Hamamatsu) with a flux of $10^{15}$ photons s$^{-1}$ cm$^{-2}$ and the final fluence of $7\times10^{18}$ photons cm$^{-2}$. The lamp has a broad spectrum from 400 to 118 nm with the main peak at 160 nm (7.7 eV) and an additional peak at 122 nm (10.2 eV) corresponding to the emission of molecular and atomic (Lyman-α line) hydrogen, respectively.

Temperature-programmed desorption (TPD) experiments were performed by heating the samples to 300 K using a linear temperature ramp of 2 K minute$^{-1}$. Infrared spectra were measured in the spectral range from 6000 to 600 cm$^{-1}$ with a resolution of 1 cm$^{-1}$ using an FTIR spectrometer (Vertex 80v, Bruker) in the transmission mode. Mass spectra during the TPD experiments were taken in the mass range from 0 to 90 amu with a scanning time of one minute for one spectrum using a quadrupole mass spectrometer (HXT300M, Hositrad).

For the presented experimental results, it is important to stress that the used atomic carbon source is characterized by a very low production of carbon clusters (<1% $C_2$ and $C_3$ molecules with respect to carbon atoms) and by the production of only ground state $C(^3P_J)$ atoms due to the formation process via thermal evaporation. This is in contrast to typically used energetic production methods, such as laser ablation of a solid target or electric discharge of a gas mixture or a liquid. In our carbon source, C atoms are assumed to be in thermal equilibrium with the tantalum tube (about 2100 K). This high temperature is required to overcome the binding energy of C atoms to tantalum and evaporate them. The amount of energy required to reach the excited singlet state $C(^1D)$ is 14 665 K (Kramida et al. 2018). Thus, the presence of C atoms in singlet states can be completely excluded.

### 3. Results

First, the reaction $C + H_2O \rightarrow H_2CO$ was studied following co-deposition of C atoms and $H_2O$ molecules at low temperatures. In Figure 1, we present the IR spectra taken after the $C + H_2O$ depositions at 10, 20, 50, and 80 K; and $C + D_2O$ depositions at 10 K. The formation of $H_2CO$ is clearly observable via its strongest vibrational bands at 1720 and 1500 $cm^{-1}$ (Watanabe & Kouchi 2002; Pirim & Krim 2014). Other, less intense bands at 2890, 2828, and 1247 $cm^{-1}$ are also observed (not shown in the figure). The main peak of doubly deuterated formaldehyde ($D_2CO$) is observable at 1670 $cm^{-1}$. Reference spectra taken after the deposition of only either carbon atoms or $H_2O$ ice show no evidence for $H_2CO$. The broad bands, in the $H_2O$ spectra around 1650 $cm^{-1}$ and in the $D_2O$ spectrum around 1500 $cm^{-1}$, are the HOH and DOD bending modes.

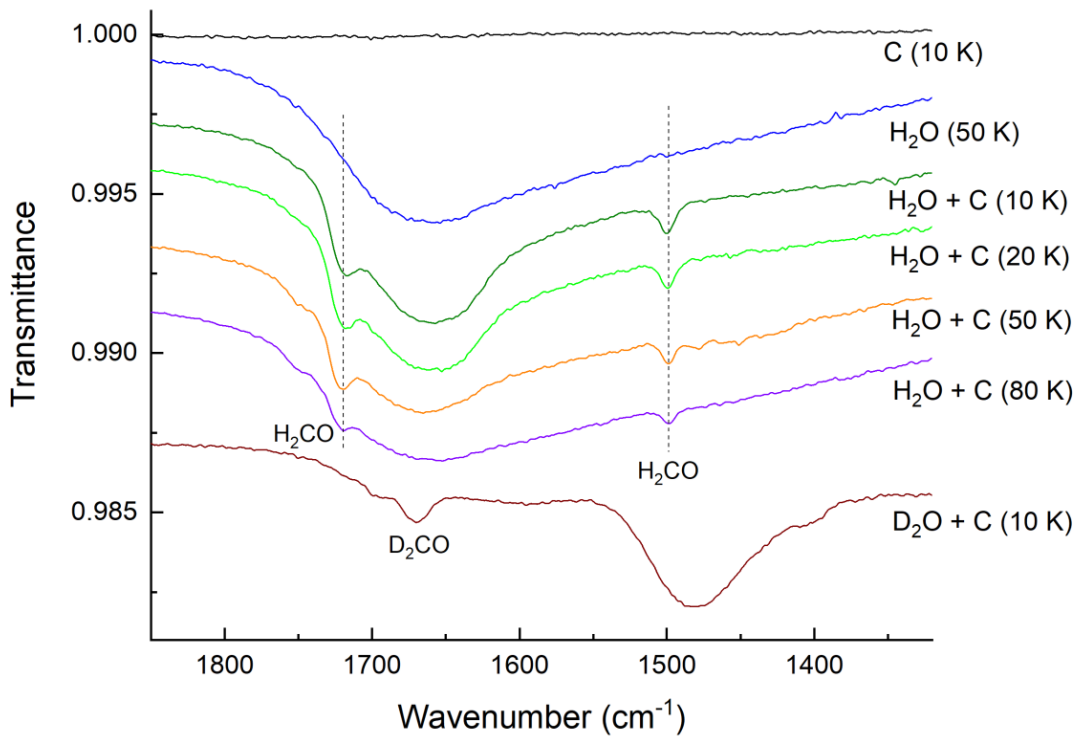

Figure 1. IR spectra taken after the depositions of (from top to bottom): C atoms; $H_2O$ ice; $H_2O$ + C at 10, 20, 50, and 80 K; and $D_2O$ + C at 10 K. The spectra are shifted vertically for clarity. Dashed lines indicate the $H_2CO$ bands.

The experiments with heavy water were performed to check if there is a difference in the measured reaction rates between the C + $H_2O$ and C + $D_2O$ reactions at the lowest temperature. As the reactions were expected to proceed via proton tunneling, a lower tunneling rate of deuterium was anticipated as it was observed for the gas phase at 50 K (Hickson, et al. 2016). However, the reaction in the gas phase was found to be pretty fast, occurring on a millisecond timescale. Our experiments showed that even at the lowest temperature of 10 K, the reaction occurs within less than one minute, the shortest time between our measurements. Thus, we could not detect any possible difference in the reaction rates of the C + $H_2O$ and C + $D_2O$ reactions.

Figure 2 presents the evolution of the column density of $H_2CO$ at 10, 20, 50 and 80 K with the deposition time. The column density values were determined from the integrated intensities of the $H_2CO$ band at 1720 cm$^{-1}$ using the band strength of 9.6x10$^{-18}$ cm molecule$^{-1}$ (Schutte,

Allamandola, & Sandford 1993). The uncertainties were determined from the maximal integrated intensity (among all the measurements) calculated for the neighborhood spectral region of the same width, which does not contain any absorption band.

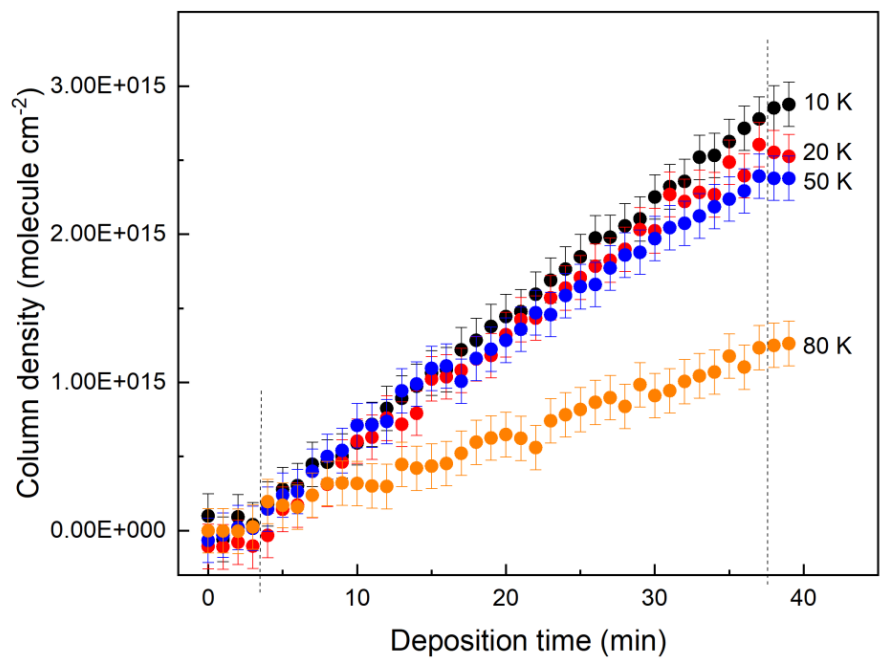

Figure 2. Evolution of the column density of $H_2CO$ at 10, 20, 50, and 80 K with the deposition time. Carbon atoms were deposited during 35 minutes between 4[th] and 38[th] minutes of the $H_2O$ deposition (marked in the figure by dashed lines).

As can be seen, the deposition of C + $H_2O$ at 10, 20, and 50 K leads to the production of similar amounts of $H_2CO$. The number of $H_2CO$ molecules is about 40 times lower compared to the number of $H_2O$ molecules, which gives a reasonable estimate for the C/$H_2O$ deposition ratio considering that the majority of the deposited carbon atoms participate in the C + $H_2O$ → $H_2CO$ reaction. As mobility and reactivity of surface species increase with the temperature, there is no reason to assume that surface formation of $H_2CO$ at 80 K proceeds less efficient as compared to lower temperatures. One of the explanations for the lower number of detected $H_2CO$ molecules at 80 K, which is close to the $H_2CO$ desorption temperature, might be chemical desorption of formaldehyde caused by dissipation of the excess reaction energy to translational energy (Fredon, Radchenko, & Cuppen 2021). However, a detailed investigation of this phenomenon is out of the scope of the present study.

Figure 3 presents TPD curves of $H_2CO$ molecules measured directly after the $H_2O$ + C deposition showing peaks at masses 30 (corresponding to $H_2CO$) and 29 (corresponding to HCO, the main fragment of $H_2CO$) with the mass(30)/mass(29) ratio of approximately 0.6 in agreement with the NIST data. Formaldehyde self-desorption around 110 K, volcano desorption due to the transformation of the water ice structure from amorphous to crystalline around 145 K and co-desorption with the $H_2O$ matrix around 160 K are clearly observable in agreement with the previous study (Noble et al. 2012).

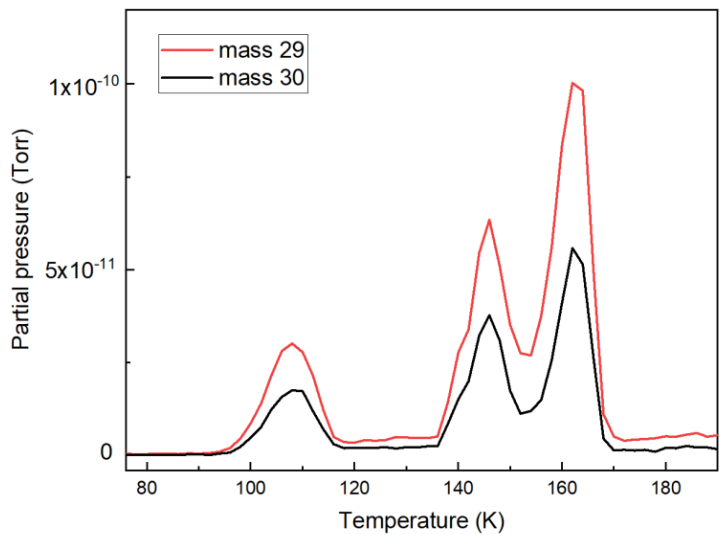

Figure 3. TPD curves for masses 29 and 30 corresponding to formaldehyde measured after the depositions of $H_2O$ + C at 10 K.

Additionally, a layered $H_2O$/Ar/C sample was deposited at 10 K, to produce thermalized and mobile carbon atoms. In this case, the amount of the produced $H_2CO$ is about five times smaller as compared to the direct C/$H_2O$ mixing. TPD curves for masses 29 and 30 corresponding to formaldehyde measured after the depositions of $H_2O$/Ar/C layers at 10 K are shown in Figure 4. The formation of formaldehyde, when C atoms and $H_2O$ molecules are initially separated by a layer of Ar, is a direct evidence of the participation of cold C atoms in the C + $H_2O$ → $H_2CO$ solid-state reaction. The smaller amount of $H_2CO$ points to the presence of the competitive reaction channel, probably, C + C$_n$. No other carbon molecules were detected by IR spectroscopy even after evaporation of all argon at about 33 K. Therefore, our results are explained by (i) efficient diffusion of C atoms in Ar ice (Thompson, et al. 1971) and (ii) the formation of amorphous carbon in addition to H$_2$CO. The latter shows that the reaction rate of C + H$_2$O is much lower compared to the rates of barrierless C + C$_n$ reactions (Thompson, et al. 1971).

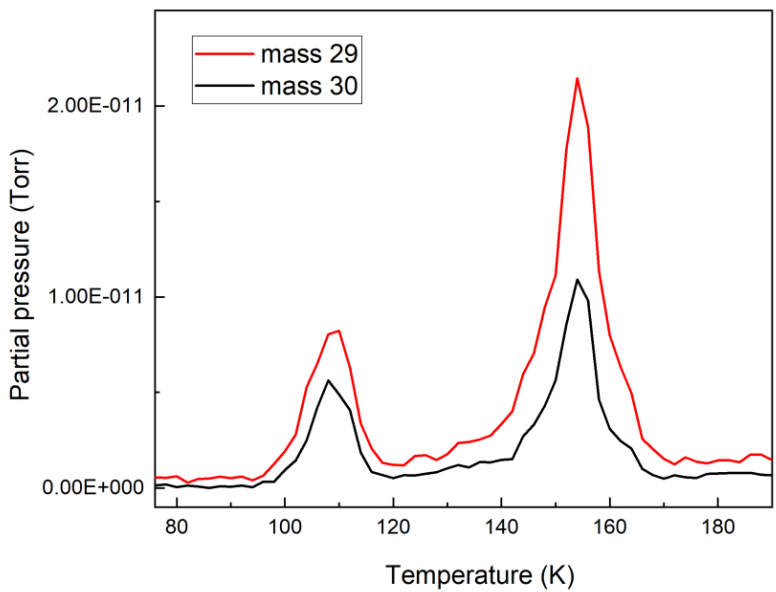

Figure 4. TPD curves for masses 29 and 30 corresponding to formaldehyde measured after the depositions of H$_2$O/Ar/C layers at 10 K.

In a second series of experiments, the C + H$_2$O ice layers were irradiated by UV photons at the corresponding deposition temperatures. Our expectation was that dissociation of H$_2$O molecules into H and OH radicals in the presence of formaldehyde will lead to the formation of more complex species, first of all, methanol (CH$_3$OH) via barrierless hydrogenation of H$_2$CO. This possibility is well-known from the experiments on hydrogenation of CO leading to the consequent formation of HCO, H$_2$CO, H$_3$CO, and CH$_3$OH (Hiraoka et al. 1994; Pirim & Krim 2014). However, UV irradiation of the ice produced by co-deposition of C and H$_2$O led to the destruction of formaldehyde with the formation of mainly CO$_2$, as illustrated in Figure 5. The amount of CO$_2$ synthesized during the irradiation at 10, 20, and 50 K (~3x10$^{15}$ molecule cm$^{-2}$) corresponds to the loss of H$_2$CO and CO (~0.3x10$^{15}$ molecule cm$^{-2}$, a contamination from the carbon source, the CO band at 2140 cm$^{-1}$ is visible in Figure 5). Only a very small amount of CH$_3$OH (~1x10$^{14}$ molecule cm$^{-2}$, about three percent of H$_2$CO) was formed as a result of the UV irradiation. The CO$_2$, CO, and CH$_3$OH ice thicknesses were calculated from their vibrational bands at 2342, 2140, and 1020 cm$^{-1}$ using the band strengths of 7.6×10$^{-17}$ (Gerakines et al. 1995), 1.1×10$^{-17}$ (Gerakines, et al. 1995) and 1.8×10$^{-17}$ cm molecule$^{-1}$ (Hudgins, et al. 1993), respectively.

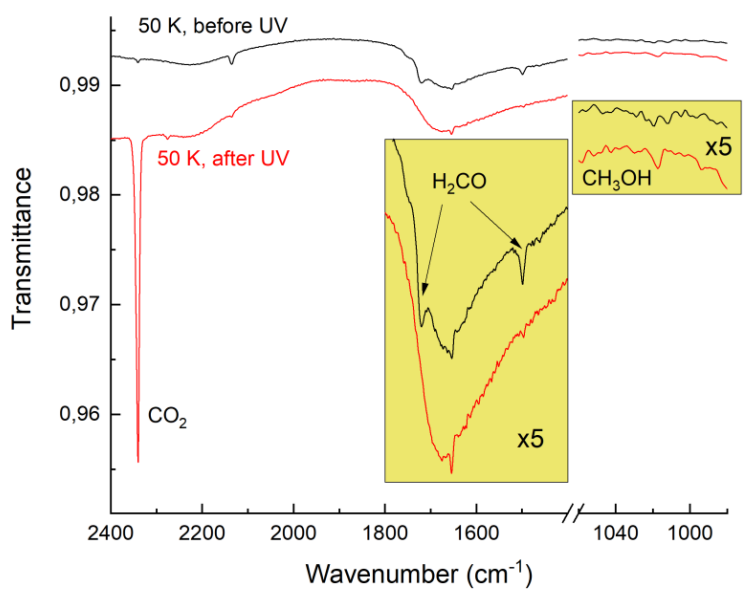

Figure 5. IR spectra taken after deposition of H$_2$O + C at 50 K before and after UV irradiation. Insets show zoomed spectra. The spectra are shifted vertically for clarity.

We cannot exclude that CO participated in the UV-triggered chemistry leading to the formation of methanol, however, the amount of CO compared to H$_2$CO was about 10 times lower, thus, the effect of the CO contamination should not be important. It is known that more complex than CH$_3$OH COMs can be formed in CO, H$_2$O, and H$_2$CO containing ices (Butscher et al. 2017; Fedoseev et al. 2017). In our experiments, with and without UV, TPD curves show very weak signals indicating desorption of different species, however, the intensity ratios at different masses does not allow us to consider unambiguous identification of other COMs by mass spectrometry. Due to the very low amounts of these species, we were also not able to detect them by IR spectroscopy.

## 4. Discussion

The C + H$_2$O reaction at $T$ = 10 K is found to be fast on the experimental timescale. The barrier of ~8.3 kcal mol$^{-1}$ (~4180 K) (Li, Xie, & Guo 2017) calculated for the entrance channel is at least twice higher than the average kinetic energies of C atoms provided by our source. Moreover, the experiment with argon provided direct evidence for spontaneous reaction C + H$_2$O → H$_2$CO at 10 K. The previous experimental and computational studies of this reaction in the gas phase showed an efficient tunneling of protons at low temperatures (Hickson, et al. 2016; Keshavarz 2019). All this indicates that in our experiments the reaction likely proceeds through tunneling when hydrogen atoms move from oxygen to carbon after the formation of the C-OH$_2$ pre-reactive complex (Hickson, et al. 2016). This is in contrast to the conclusion on non-reactivity of ground state C atoms with water molecules at low temperatures (Schreiner & Reisenauer 2006), where the C + H$_2$O → H$_2$CO reaction was observed in Ar matrixes, but was assigned to the presence of singlet C atoms. We note that Schreiner & Reisenauer worked with "hot" C atoms, and our experiments show that also a low-temperature reaction channel exists.

In contrast to the gas phase study (Hickson, et al. 2016), formaldehyde molecules formed in the solid state are not dissociated and are stabilized by transferring the excess reaction energy to a third body (the ice surface). The need to tunnel through the barrier implies a rather low reaction rate compared to rates of barrierless reactions. Therefore, if competitive barrierless reaction channels are available, the reaction would likely follow this barrierless route as it was observed in our experiment with Ar, where a smaller amount of H$_2$CO was detected. It would mean that in astrophysical environments when C atoms accrete on the surface of dust grains, the formation of H$_2$CO is efficient if the accretion site contains water molecules only. In the presence of CO or NH$_3$ molecules on the accretion sites, the barrierless formation of C$_2$O or H$_2$CNH molecules could be more efficient (Krasnokutski, et al. 2020).

Formaldehyde is among the more abundant molecules in the solid state in cold astrophysical environments (Boogert, Gerakines, & Whittet 2015). The formation of formaldehyde is one of the important steps towards molecular complexity. First, barrierless hydrogenation of formaldehyde leads to the formation of methanol, which, in turn, is well-known as a starting point for the formation of more complex organic molecules (Oberg et al. 2009; Vasyunin & Herbst 2013; Balucani, Ceccarelli, & Taquet 2015). Second, the thermal formation of methyleneglycol (HOCH$_2$OH), a product of the thermal reaction between H$_2$CO and H$_2$O, and formaldehyde polymers (polyoxymethylene (POM), HO–(CH$_2$–O)$_n$–H)) in H$_2$O:NH$_3$:H$_2$CO ice mixtures, where NH$_3$ played the role of a catalyst, was clearly demonstrated (Duvernay et al. 2014). However, the formation of methyleneglycol and POM occurs without NH$_3$ as well, just on a longer timescale (Patrice Theulé, private communication). In cold astrophysical environments, such as molecular clouds, protostars and planet-forming disks beyond the snowline, both factors, time and NH$_3$ are clearly available. Third, the studied reaction also opens a straight forward way to sugar formation. It was shown that H$_2$CO reacts with its isomer HCOH to give glycolaldehyde (HOCH$_2$-CHO) in a nearly barrierless reaction (Eckhardt et al. 2018). With further addition of HCOH molecules, the formation of glyceraldehyde is possible. The HCOH radical is an intermediate in the studied C + H$_2$O reaction. The transformation of HCOH to formaldehyde was also found to take place due to the tunnelling of hydrogen atoms (Schreiner et al. 2008). Therefore, if the reaction C + H$_2$O would take place close to another H$_2$CO molecule, the reaction HCOH + H$_2$CO → HOCH$_2$-CHO should be expected.

The typically considered mechanism of the H$_2$CO formation, hydrogenation of CO ice, is relevant to the regions beyond the CO snowline (around 30 K). The C + H$_2$O reaction shows an alternative pathway to H$_2$CO. In warmer environments, particularly, in the inner regions of planet-forming disks, CO is mainly in the gas phase, however, H$_2$O ice is present at much higher temperatures (until 140 – 170 K depending on the disk model). As mobility and reactivity of surface species increase with the temperature, there is no reason to assume that surface formation of H$_2$CO at higher temperatures proceeds less efficient as compared to lower temperatures. Thus, considering the presence of C atoms, the C + H$_2$O reaction might be an important source of formaldehyde in planet-forming disks. The reaction should be included into existing models to evaluate its importance for astrochemical networks in different astrophysical environments.

To date, most of the work on the formation of complex organic and prebiotic molecules under astrophysically relevant conditions was done through the energetic processing of molecular ices (UV irradiation or ion/electron/proton bombardment). Here, we refer the reader to a couple of review papers (Öberg 2016; Arumainayagam et al. 2019). However, in the early evolutionary stages of interstellar clouds, the main state of matter is an atomic gas. The most important atoms for prebiotic astrochemistry (such as C, H, O, N, S, P, and etc.) are radicals. These radicals may react and build large organic molecules without an additional energy input as it was shown in recent papers on the "non-energetic" formation of glycine (Krasnokutski, et al. 2020; Ioppolo, et al. 2021). During the formation of dense and cold regions (molecular clouds) in the ISM, the condensation of atoms on the surface of dust grains should lead to the synthesis of complex organic species.

In our experiments, UV irradiation of the ices produced by co-deposition of C and $H_2O$ reactants played mainly a destructive role (with respect to the formation of more or less complex species as compared to $H_2CO$). The conversion of $H_2CO$ into $CO_2$ rather than into $CH_3OH$ was observed. However, using the qualitative approach typical for many previous works on the formation of astrophysically relevant organic molecules in ices by UV irradiation (neglecting the amounts of the formed and destroyed molecules), we could conclude that UV light plays a constructive role leading to the formation of more complex species (methanol). On the other hand, if quantities are considered, our experiments show that the UV radiation plays mainly a destructive role. Only about a few percent of $H_2CO$ was converted into a more complex species (methanol) and the rest was converted into a simpler species ($CO_2$). This result challenges the role of energetic processing in the synthesis of complex organic molecules under astrophysically relevant conditions and might be particularly important for the case of the formation of prebiotic molecules, which are commonly rather fragile. Therefore, the formation of organics and prebiotic molecules could be most efficient in areas where there are less energetic triggers of reactions, such as UV photons or cosmic rays. However, more studies are required for a better understanding of the role of energetic and "non-energetic" processes in the formation of complex organic and prebiotic molecules.

To conclude, the solid-state formation of $H_2CO$ at temperatures relevant to the ISM and colder regions of circumstellar environments of young stars proceeds through the C + $H_2O$ reaction involving quantum tunneling mechanism. This alternative, "non-energetic" route to formaldehyde has to complement astrochemical models and may lead to better understanding of the processes leading to the formation of organics in space.

**Acknowledgments**

AP and CJ acknowledge support from the Research Unit FOR 2285 "Debris Disks in Planetary Systems" of the Deutsche Forschungsgemeinschaft (grant JA 2107/3-2). SK acknowledges support from the Deutsche Forschungsgemeinschaft (grant KR 3995/4-1). TH acknowledges support from the European Research Council under the Horizon 2020 Framework Program via the ERC Advanced Grant Origins 83 24 28.acknowledges support from the European Research Council under the Horizon 2020 Framework Program via the ERC Advanced Grant Origins 83 24 28.